\documentclass[conference]{IEEEtran}

\usepackage{algorithmic}
\usepackage{algorithm}

\ifCLASSINFOpdf

\else

\fi

\usepackage{graphicx} 
\usepackage{footnote}

\usepackage[letterpaper, left=0.63in, right=0.63in, top=1in, bottom=2.7cm]{geometry}

\usepackage{xcolor}
\usepackage{soul}

\begin{document}

\title{Performance Analysis of Sub-band Full-duplex Cell-free Massive MIMO JCAS Systems}

\author{\IEEEauthorblockN{Kwadwo Mensah Obeng Afrane, Yang Miao, and André B.J. Kokkeler}
\IEEEauthorblockA{Faculty of EEMCS, University of Twente\\
Enschede, The Netherlands\\
k.m.obengafrane@utwente.nl, y.miao@utwente.nl, a.b.j.kokkeler@utwente.nl}}

\maketitle

\begin{abstract}

In-band Full-duplex joint communication and sensing systems require self interference cancellation as well as decoupling of the mutual interference between UL communication signals and radar echoes.
We present sub-band full-duplex as an alternative duplexing scheme to achieve simultaneous uplink communication and target parameter estimation in a cell-free massive MIMO system. Sub-band full-duplex allows uplink and downlink transmissions simultaneously on non-overlapping frequency resources via explicitly defined uplink and downlink sub-bands in each timeslot. Thus, we propose a sub-band full-duplex cell-free massive MIMO system with active downlink sensing on downlink sub-bands and uplink communication on uplink sub-band. In the proposed system, the target illumination signal is transmitted on the downlink (radar) sub-band whereas uplink users transmit on the uplink (communication) sub-band. By assuming efficient suppression of inter-sub-band interference between radar and communication sub-bands, uplink communication and radar signals can be efficiently processed without mutual interference. We show that each AP can estimate sensing parameters with high accuracy in SBFD cell-free massive MIMO JCAS  systems.

\end{abstract}

\IEEEpeerreviewmaketitle

\section{ Introduction}

\begin{figure}[!t]
\centering

\includegraphics[width=2.1in]{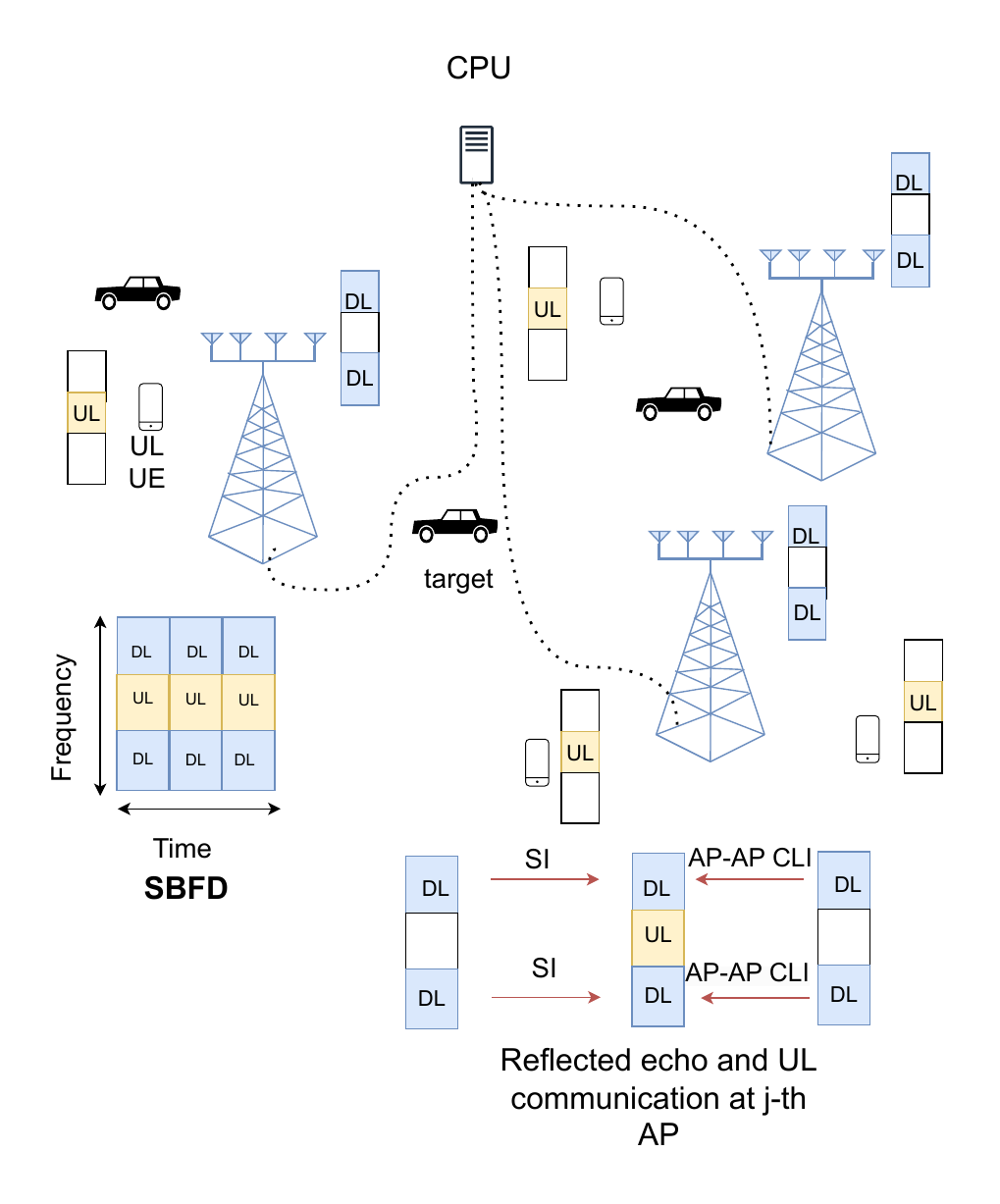}

\caption{Illustrating the concept of SBFD CF massive MIMO system}
\label{fig1}
\end{figure}

The premise for proposing sub-band full-duplex (SBFD) by \cite{ji_extending_2021} under the name of extended division duplex (XDD) was to address the shortcomings of traditional time division duplex (TDD) and in-band full-duplex (IBFD). In TDD systems, the assignment of most of the time resources for downlink (DL) transmission results in smaller uplink (UL) coverage. IBFD on the other hand also introduces self-interference (SI), access point (AP)-to-AP and user equipment(UE)-to-UE interference which requires advanced interference mitigation techniques to suppress. SBFD enables simultaneous UL and DL transmissions on non-overlapping frequency resources within the same unpaired TDD carrier \cite{ji_extending_2021}. Unlike frequency division duplex (FDD) which employs separate fixed UL and DL frequency bands with a large  guard-band (GB), SBFD utilizes a single frequency band for UL and DL which can be assymetrically assigned \cite{mokhtari_modeling_2023}. In the context of simultaneous communication and monostatic sensing, joint transceiver design may require, in addition to SI cancellation, decoupling of the mutual interference between UL communication signals and radar echoes. The authors in \cite{temiz_dual-function_2022} proposed to first detect and demodulate UE signals and subsequently subtract from received signal to obtain radar returns. However in such a system, a decoding error in the communication signal processing can lead to error propagation which degrades the radar performance. Thus in this work we adapt the semi-static SBFD frame \cite{38.858} structure for a cell-free (CF) massive MIMO joint communication and sensing (JCAS) system as shown Fig.~\ref{fig1}. The sub-band (SB) location in the frequency domain is fixed and explicitly defined for all time slots and all APs share the same centrally defined frame structure at the central processing unit (CPU). Compared to the existing SBFD specification, which is defined for only cellular communications systems, We propose to utilize DL SBs for active DL sensing and a defined non-overlapping UL SBs for UL communications. We show that sensing parameter estimates can be accurately determined in SBFD JCAS systems.  Beyond decoupled UL and radar reception on non-overlapping orthogonal subcarriers, SBFD JCAS imposes less strict requirements on SI mitigation at each AP for UL communication. Furthermore, the distributed and cooperative nature of CF MIMO systems ensures uniform quality of service for communication users and also enhances localization and tracking accuracy. Additionally, the CPU, can globally define task-dependent SBFD configurations across APs to meet varying sensing and communication requirements.

\section{SYSTEM MODEL}
As shown in Fig.~\ref{fig1}, we investigate CF massive MIMO system with semi-static SBFD APs - UL and DL SB locations are explicitly configured across all time slots and similarly defined for all APs. We use the "DUD" SBFD configuration \cite{38.858} which implies that each SBFD slot consists of one UL SB at the center of the channel bandwidth and two DL SBs at both ends \ref{fig1}.
Additionally, GBs are placed between UL and DL SBs to improve isolation between both SBs. In the proposed system, all APs transmit the same OFDM waveforms in the DL SB to illuminate $T$ targets via the conjugate sensing beamforming in target direction obtained via previous estimation.
Simultaneously, SBFD-aware half-duplex UEs transmit UL communication symbols occupied in the UL SB. Since each AP performs both communication and sensing simultaneously both UL communication and echoes may be received at the same time. For ease of exposition the presented numerical example ignores inter-SB SI and cross link interference (CLI) which may occur as a result of transceiver imperfections. It must however be noted that radar reflections are still impaired by intra-SB SI and intra-SB CLI since the radar subsystem still operates in IBFD mode. Both interferences are modeled as residuals after SI cancellation techniques in the spatial, analog and/or digital domain. 
Furthermore, all $J$ APs operate in monostatic mode thus multi-static echo reflections from other APs are not considered. Radar-interference free UL communications signals are processed with the maximum ratio combining (MRC)  receive beamformer in a distributed manner and linearly combined at the CPU.
\section{Numerical Example}
\begin{figure}[!t]
\centering
\includegraphics[width=2.2in]{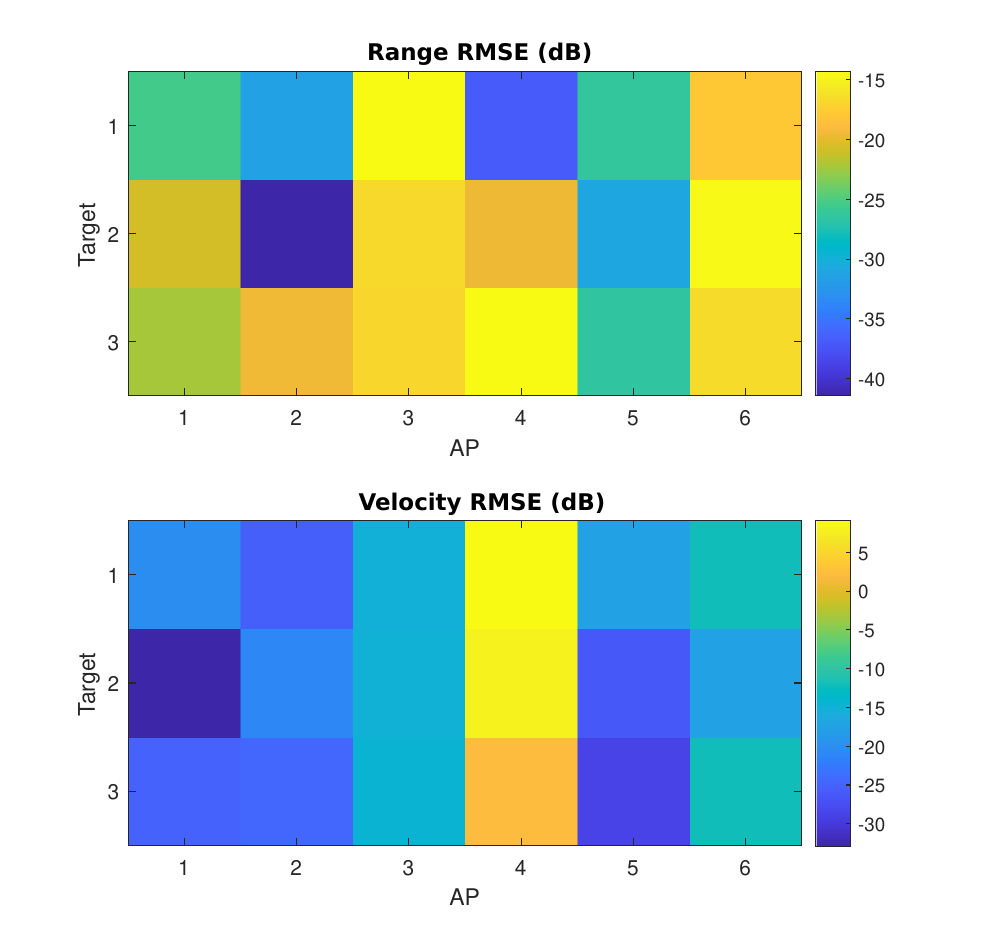}
\caption{RMSE Performance for $J=6$, $T=3$}
\label{fig3}
\end{figure}

\begin{figure}[!t]
\centering
\includegraphics[width=2.2in]{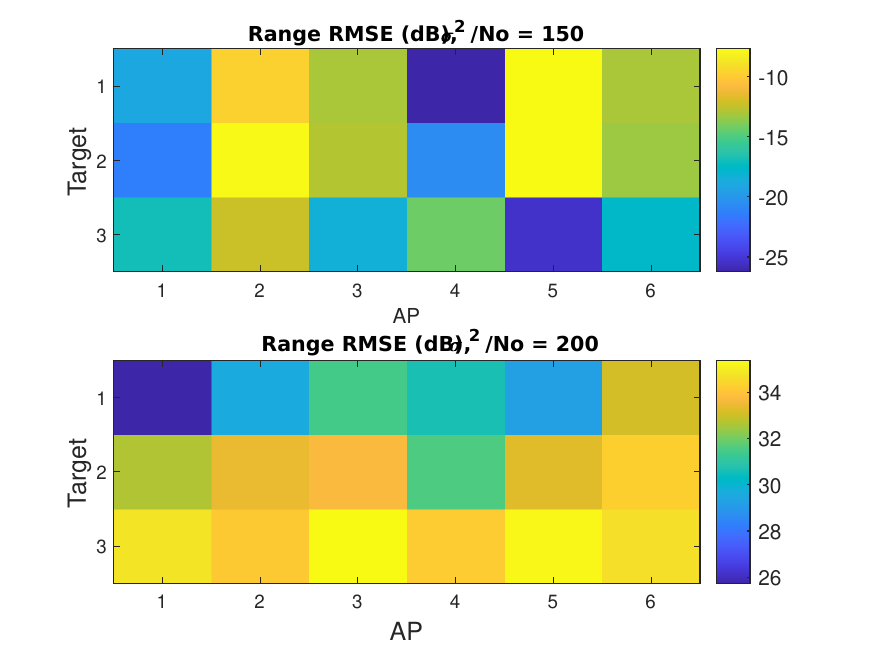}
\caption{RMSE Performance for different residual SI values $J=6$, $T=3$}
\label{fig4}
\end{figure}

To analyze the performance of the system we use the estimation of signal parameters via rotational invariance techniques (ESPRIT) to estimate the range and radial velocity of the target at each AP. This is a subspace-based method for the estimation of line spectra and was chosen due to its simplicity and suitability for OFDM radar \cite{Braun2014_1000038892}. The following system parameters were used; A $250 \times 250 m^2$ area with $6$ APs and $5$ non-mobile single antenna UL UEs. The setup includes $3$ targets moving at velocities, $(18,28 m\backslash s),(10,-28 m\backslash s)$ and $(21,26 m\backslash s)$ respectively. All APs are equipped with $4$ half-wavelength-spaced antennas deployed as a ULA. We consider received target reflections and UL communication at an arbitrary SBFD time slot with 14 OFDM symbols, at 7 GHz, a subcarrier spacing of 30kHz and a bandwidth of  50 MHz.Out of 133 resource blocks (RB), DL,UL and GB occupy 100, 27 and 6 RB's respectively with configuration $\{\frac{DL}{2}, \frac{GB}{2}, UL, \frac{GB}{2} \frac{DL}{2}\}$. A residual SI/CLI power $(\sigma ^2/N_o)$ of $100$ dB is and an SNR of $10$ dB is assumed.
Fig. \ref{fig3} shows the root mean squared error (RMSE) of target range and velocity estimates at each AP. It is evident that all APs can accurately estimate the range and velocity of all targets. It is observed that the variation in RMSE for different APs with respect to the same target can be associated to the difference in the AP-target distances, as well as the varying levels of target-free CLI interference with respect to AP-AP distance. However, in SBFD CF massive MIMO, the sensing parameter estimation performance is impacted by the SI/CLI interference cancellation capability of the system as shown in Fig. \ref{fig4} As such doubling the residual interference results in a more than 300 percent increase in the range RMSE.

\section{Conclusion}
We propose a SBFD CF massive MIMO system with active DL sensing on DL SB and UL communication on UL SB.
The numerical results demonstrate that the each AP can estimate sensing parameters with high accuracy in SBFD CF massive MIMO JCAS  systems. However, even though interference between UL communication and radar signal is mitigated by the SBFD operation, the performance of the radar subsystem is limited by the SI/CLI cancellation capability of the system. Doubling the residual SI/CLI power results in a more than 300 percent increase in the range RMSE. Future work would consider multi-static sensing, transceiver imperfections well as inter-SB interference between UL and DL SBs.

\bibliographystyle{ieeetr}

\bibliography{EuCnC}

\begin{thebibliography}{1}

\bibitem{ji_extending_2021}
H.~Ji {\em et~al.}, ``Extending 5g {TDD} coverage with {XDD}: Cross division duplex,'' {\em {IEEE} Access}, vol.~9, pp.~51380--51392.

\bibitem{mokhtari_modeling_2023}
M.~Mokhtari {\em et~al.}, ``Modeling and system-level performance evaluation of sub-band full duplexing for 5g-advanced,'' {\em {IEEE} Access}, vol.~11, pp.~71503--71516.

\bibitem{temiz_dual-function_2022}
M.~Temiz {\em et~al.}, ``A dual-function massive {MIMO} uplink {OFDM} communication and radar architecture,'' {\em {IEEE} Trans. Cogn. Commun. Netw.}, vol.~8, no.~2, pp.~750--762.

\bibitem{38.858}
3GPP, ``3rd generation partnership project; technical specification group radio access network; study on evolution of nr duplex operation (release 18),'' Tech. Rep. TR 38.858 V18.1.0 (2024-03), 3GPP, 2024.

\bibitem{Braun2014_1000038892}
K.~M. Braun, {\em OFDM Radar Algorithms in Mobile Communication Networks}.
\newblock PhD thesis, 2014.

\end{thebibliography}
\end{document}